\documentstyle[epsf]{mn}

\newif\ifAMStwofonts

\def\simlt{\lower.5ex\hbox{$\; \buildrel < \over \sim \;$}}
\def\simgt{\lower.5ex\hbox{$\; \buildrel > \over \sim \;$}}

\def\be{\begin{equation}}
\def\ee{\end{equation}}

\def\bee{\begin{eqnarray}}
\def\eee{\end{eqnarray}}

\ifoldfss
  \ifCUPmtlplainloaded \else
    \NewTextAlphabet{textbfit} {cmbxti10} {}
    \NewTextAlphabet{textbfss} {cmssbx10} {}
    \NewMathAlphabet{mathbfit} {cmbxti10} {} 
    \NewMathAlphabet{mathbfss} {cmssbx10} {} 
  \fi
  \ifAMStwofonts
    \ifCUPmtlplainloaded \else
      \NewSymbolFont{upmath} {eurm10}
      \NewSymbolFont{AMSa} {msam10}
      \NewMathSymbol{\upi}     {0}{upmath}{19}
      \NewMathSymbol{\umu}     {0}{upmath}{16}
      \NewMathSymbol{\upartial}{0}{upmath}{40}
      \NewMathSymbol{\leqslant}{3}{AMSa}{36}
      \NewMathSymbol{\geqslant}{3}{AMSa}{3E}

      \let\leq=\leqslant 
      \let\geq=\geqslant 
    \fi
  \fi
\fi 

\ifnfssone
  \newmathalphabet{\mathit}
  \addtoversion{normal}{\mathit}{cmr}{m}{it}
  \addtoversion{bold}{\mathit}{cmr}{bx}{it}
  \newmathalphabet{\mathbfit} 
  \addtoversion{normal}{\mathbfit}{cmr}{bx}{it}
  \addtoversion{bold}{\mathbfit}{cmr}{bx}{it}
  \newmathalphabet{\mathbfss} 
  \addtoversion{normal}{\mathbfss}{cmss}{bx}{n}
  \addtoversion{bold}{\mathbfss}{cmss}{bx}{n}
  \ifAMStwofonts
    \ifCUPmtlplainloaded \else
      \UseAMStwoboldmath
      \makeatletter
      \new@mathgroup\upmath@group
      \define@mathgroup\mv@normal\upmath@group{eur}{m}{n}
      \define@mathgroup\mv@bold\upmath@group{eur}{b}{n}
      \edef\UPM{\hexnumber\upmath@group}
      \new@mathgroup\amsa@group
      \define@mathgroup\mv@normal\amsa@group{msa}{m}{n}
      \define@mathgroup\mv@bold\amsa@group{msa}{m}{n}
      \edef\AMSa{\hexnumber\amsa@group}
      \makeatother
      \mathchardef\upi="0\UPM19
      \mathchardef\umu="0\UPM16
      \mathchardef\upartial="0\UPM40
      \mathchardef\leqslant="3\AMSa36
      \mathchardef\geqslant="3\AMSa3E

      \let\leq=\leqslant 
      \let\geq=\geqslant 
    \fi
  \fi
\fi 

\ifnfsstwo
  \DeclareMathAlphabet{\mathbfit}{OT1}{cmr}{bx}{it}
  \SetMathAlphabet\mathbfit{bold}{OT1}{cmr}{bx}{it}
  \DeclareMathAlphabet{\mathbfss}{OT1}{cmss}{bx}{n}
  \SetMathAlphabet\mathbfss{bold}{OT1}{cmss}{bx}{n}
  \ifAMStwofonts
    \ifCUPmtlplainloaded \else
      \DeclareSymbolFont{UPM}{U}{eur}{m}{n}
      \SetSymbolFont{UPM}{bold}{U}{eur}{b}{n}
      \DeclareSymbolFont{AMSa}{U}{msa}{m}{n}
      \DeclareMathSymbol{\upi}{0}{UPM}{"19}
      \DeclareMathSymbol{\umu}{0}{UPM}{"16}
      \DeclareMathSymbol{\upartial}{0}{UPM}{"40}
      \DeclareMathSymbol{\leqslant}{3}{AMSa}{"36}
      \DeclareMathSymbol{\geqslant}{3}{AMSa}{"3E}

      \let\leq=\leqslant 
      \let\geq=\geqslant 
    \fi
  \fi
\fi 

\ifCUPmtlplainloaded \else
  \ifAMStwofonts \else 
    \def\upi{\pi}
    \def\umu{\mu}
    \def\upartial{\partial}
  \fi
\fi

\title[DM profiles and the Jeans equation]
{Dark matter density profiles from the Jeans equation}
\author[Steen H. Hansen]
{Steen H. Hansen\\
University of Zurich, Winterthurerstrasse 190,
CH-8057 Zurich, Switzerland }
\date{Draft version \today}
\pagerange{\pageref{firstpage}--\pageref{lastpage}}
\pubyear{2004}

\begin{document}

\maketitle

\label{firstpage}

\begin{abstract}
We make a simple analytical study of radial profiles of dark matter
structures, with special attention to the question of the central
radial density profile.  We let our theoretical assumptions be guided
by results from numerical simulations, and show that at any radius
where both the radial density profile, $\rho$, and the
phase-space-like density profile, $\rho/\sigma^{\epsilon}$, are exact
power laws, the only allowed density slopes in agreement with the
spherical symmetric and isotropic Jeans equation are in the range $1
\leq \beta \leq 3$, where $\beta \equiv - d {\rm ln}\rho/d {\rm
ln}r$. We also allow for a radial variation of these power laws, as
well as anisotropy, and show how this allows for more shallow central
slopes.
\end{abstract}




\section{Introduction}
The formation and evolution of Dark Matter (DM) structures is in
principle very simple since it only involves gravity.  Despite this
fact, density profiles of dark matter halos have become one of the
most challenging issues for our understanding of cold dark matter
structure formation. Numerical simulations provide predictions of
steep central density cusps with power law slopes, $\rho \sim
r^{-\beta}$, with $\beta$ from $1$ to $1.5$ within a few percent of
the virial radius of the halo~\cite{nfw,moore}.  Recent careful
studies~\cite{diemand,reed,navarro} indicate that the resolved region
has still not converged on a central density slope, so in principle
the central power slope may be even shallower.

The steep inner numerically resolved slopes are, however, not
supported by observations.  By measuring the rotation curve of a
galaxy one can in principle determine the density profile of its DM
halo. Low surface brightness galaxies and spirals, where the observed
dynamics should be DM dominated, seem to show slowly rising rotation
curves
\cite{rubin85,courteau97,palunas00,blok01,blok02,salucci01,swaters02,corb03}
indicating that these DM halos have constant density cores.  Galaxy
clusters, where baryons can play even less of a role, may show a
similar discrepancy. Arcs \cite{sand02} and strong lensing fits of
multiple image configurations and brightnesses \cite{tyson98} also
indicate shallow cores in clusters.  All these observations could be
in agreement with N-body simulations only if either the very central
region is really not cuspy, or if cusps could somehow be erased during
galaxy formation.

It is therefore very important to understand if the pure dark matter
central density slopes can really be as steep as indicated by
numerical simulations, in order to understand if one needs to invoke
baryonic physics to reach agreement with observations.
Baryonic structures are often observed to to have central cores, which
is possibly even understood theoretically~\cite{hansenstadel}.

Several attempts have been made for an analytical derivation of the
density profile~\cite{bertschinger,syer,subra,hiotelis,dekel,alberto},
and none seem to present a clear and simple explanation for the
findings of N-body codes.  We will not attempt to answer the very
difficult question of the actual formation of DM structures here, but
will instead simply consider the Jeans equation and ask {\em which}
equilibrium DM structures are in agreement with the spherically
symmetric Jeans equation. We will be guided by the findings of
numerical simulations and only consider the special cases where the
phase-space-like density, $\rho/\sigma^{\epsilon}$, is a power law in
radius for some positive $\epsilon$.

The normal use of the Jeans equation for collisionless systems
\cite{hernquist90,tremaine94} is to assume a given radial density
profile, $\rho(r)$, and then solve the Jeans equation to get the
corresponding velocity dispersion, $\sigma^2(r)$.  This can be done
analytically for sufficiently nice density profiles, and can always be
done numerically. The basic result is that the Jeans equation can allow
for almost {\em any} shape of the density profile.
An alternative approach is instead to assume the
form of the phase-space density, $\rho/\sigma^3(r)$, and then solve the Jeans
equation to get the corresponding density profile
\cite{taylor01}. Also this can always be done numerically, and even
analytically for sufficiently nicely behaving velocity dispersions.

We will show below that for sufficiently simple phase-space (like)
densities, this approach can provide analytical insight into the
allowed range of density profiles.  One example hereof is that if both
the central density profile and the phase-space-like density are {\em
exact} power laws, then the central density profile of an isotropic DM
structure cannot be more shallow than an NFW profile with $\beta =1$.

\section{Exact power laws}
Let us first consider the case where the 
coarse grained radial density profile is an
exact power law
\begin{equation}
\rho \sim r^{-\beta} \, ,
\end{equation}
at a given radius. Now, Taylor \& Navarro~\shortcite{taylor01}
observed that the phase-space density, $\rho/\sigma^3$, from N-body
simulations approximately follows a power law.  Recent high resolution
N-body results confirm that this is approximately correct in the
equilibrated inner region, where substructures are unimportant
(Diemand, private communication).
\footnote{Taylor \& Navarro~\shortcite{taylor01}
considered  spherical bin averages of $\rho(r)$ and $\sigma^3(r)$, and then
took the ratio, $\rho/\sigma^3$. The actual phase-space
density, which is spherical averages of $\rho/\sigma^3$, differs due to
substructures~\cite{diemand2,arad}. We consider only the equilibrated
inner region of the DM structure where there is no difference.}
We will here make a slightly weaker
assumption, namely that a phase-space-{\em like} density  profile is an
exact power law in the very central region
\begin{equation}
\frac{\rho}{\sigma^\epsilon} \sim  r^{-\alpha} \, ,
\label{eq:phase}
\end{equation}
with unknown real numbers $\epsilon>0$ and $\alpha$.  
Taylor \& Navarro~\shortcite{taylor01} found $\epsilon=3$ and
$\alpha=1.875$.
Recent high
resolution N-body simulations do infact support this assumption for
the very central numerically resolved region with $\epsilon$ of the
order 2-3. We will here not attempt to understand why the
phase-space-like density is a power law in the centrally resolved
region. A reason therefore must be sought at a deeper level, maybe
through a solution to the collisionless Boltzmann equation.

For spherically symmetric and isotropic systems the Jeans
equation can be written \cite{binneytremaine,taylor01}
through the use of Poisson equation (for a self gravitating system)
\begin{equation}
\frac{d}{dr} \left( \frac{-r^2}{G\rho}  \frac{d}{dr} 
\left(\sigma^2 \rho \right) \right) = 4\pi \rho r^2 \, .
\label{eq:jeans}
\end{equation}
This assumption of a spherical, isotropic system is guided by the
numerical N-body results in the central part of the DM structure.  The
inclusion of anisotropy, e.g. $A_\beta = 1 - {\overline
{v_\theta^2}}/{\overline {v_r^2}}$, gives another term in the Jeans
equation, $2 A_\beta d(r\sigma^2)/dr$, and will therefore increase the
space of solutions~\footnote{We use an unusual notation for the
anisotropy parameter, $A_\beta$, to avoid confusion with the $\beta$
in the density profile.}, and we will later show how. We will leave
non-spherical structures for a later analysis.

Under the assumption of power laws, eq.~(\ref{eq:jeans}) can now be
written
\begin{equation}
-C_1 \, C_2 \, r^{2(\alpha-\beta)/\epsilon} = C_3 \, r^{2-\beta}  \, ,
\label{eq:jeanscc}
\end{equation}
where the two coefficients $C_1$ and $C_2$ come from the radial
differentiations, e.g. $C_1 = d{\rm ln} (\sigma^2 \rho)/d {\rm ln} r$,
and the last coefficient $C_3$ is a positive constant.
Clearly, the radial power-laws in eq.~(\ref{eq:jeanscc})
have to agree, giving
\begin{equation}
\beta = \frac{2 \left( \epsilon - \alpha \right) }{\left( 
\epsilon -2 \right) } 
\, .
\label{eq:beta}
\end{equation}
Moreover, for the Jeans equation, eq.~(\ref{eq:jeans}), to make sense, the
product $C_1 \, C_2$ must be negative. If the product, $C_1 \, C_2$, is
positive, then there will be something negative on the lhs of
eq.~(\ref{eq:jeanscc}) and something positive on the rhs.
We can thus find the points where the Jeans equation breaks down
by solving $C_1 \, C_2 = 0$.
This is a simple
quadratic equation in $\alpha$, with solutions
\begin{equation}
\alpha = 2 \pm \left(\epsilon/2 - 1 \right) \, .
\label{eq:alpha}
\end{equation}
That is, when $\alpha$ has the value in eq.~(\ref{eq:alpha}) then
the lhs of the Jeans equation is zero.

Combining the two results in eqs.~(\ref{eq:beta}, \ref{eq:alpha}) tell
us, that the {\em only} allowed values for the density slope are in
the range
\begin{equation}
1 \leq \beta \leq 3 \, ,
\label{eq:betarange}
\end{equation} 
and one thus concludes that for this most simple case of pure power
laws, the central density profile of pure dark matter structures
cannot be more shallow than $\rho \sim r^{-1}$. Please note that this
result is obtained for rather general power laws like
eq.~(\ref{eq:phase}) with {\em any} value of $\epsilon$ and $\alpha$.  The
results of recent N-body simulations tell us that locally the density
profile can be approximated by a power law, and furthermore one can
always find a value for $\epsilon$ such that eq.~(\ref{eq:phase})
holds true locally. Therefore, for any radius in the resolved region
of N-body simulations where substructure is not important, the density
profile must be in the range of eq.~(\ref{eq:betarange}).

When one includes non-isotropic systems, where $A_\beta \neq 0$, then
one finds the lower limit to be
\begin{equation}
\beta_{\rm min} = 1 + A_\beta \, ,
\label{eq:anisotropy}
\end{equation}
which implies
that for a negative $A_\beta$ one can have more shallow profiles,
e.g. a core in density for sufficiently circular orbits.
For purely radial motion the most shallow profile is $\beta_{\rm min}
=2$, which is in agreement with  numerical findings for the
spherical infall model~\cite{lokas}.
In general $A_\beta$ can take any value in the range, $-\infty < A_\beta < 1$
\cite{mamonlokas}, naturally constrained by $\beta$ being always 
non-negative~\cite{tremaine94,hansen04},
however, it should be kept in mind that 
numerical simulations find very little anisotropy in the very 
central region, $A_\beta \approx 0$ \cite{moorePRD}.
The central isotropy can also be understood from a fundamental
statistical mechanics point of view~\cite{hansen04}.

\section{Almost power laws}
If the density profile and the phase-space-like density profile are
not exact power laws, then our findings in eq.~(\ref{eq:betarange}) 
may potentially be invalid. In order
to investigate this question we can make expansions around power laws.
As we will see, this will also indicate how far beyond the resolved 
region it makes sense to extend our findings. We therefore write
\begin{eqnarray}
\rho &\sim& r^{- \beta (r)} \, , \\
\frac{\rho}{\sigma^{\epsilon} } & \sim & r^{- \alpha (r)} \, ,
\label{eq:phasespace}
\end{eqnarray}
where $\beta(r)$ and $\alpha(r)$ are now slowly varying functions of
radius.  We choose a fixed, radially-independent $\epsilon$.  Now, let
us make a Taylor expansion around the radius $r_{-1}$, where $\beta =
1$, using $\beta ' \equiv d \beta / d{\rm ln } r$ and $\alpha ' \equiv
d \alpha/ d{\rm ln } r$.  It should therefore be kept in mind that
this Taylor expansion only holds sufficiently nearby the point of
expansion, such that the higher derivative can be ignored, $\beta '
\gg \beta '' \, {\rm ln}r$.  The Jeans equation again looks like
eq.~(\ref{eq:jeanscc}), and the radial powers again lead to the
expression for $\beta$ in eq.~(\ref{eq:beta}). However, the
coefficients are now different
\begin{eqnarray}
C_1 &=& -\beta + \frac{2\left( \alpha - \beta \right) }{\epsilon} +
{\rm ln}(r/r_{-1}) \, 
\frac{2}{\epsilon } \left( \alpha ' - \beta ' \left( 1 +\epsilon /2 
\right) \right) \nonumber \\
C_2 &=& 1 + \frac{2 \left( \alpha -\beta \right) }{\epsilon}
+{\rm ln} (r/r_{-1}) \, \frac{2}{\epsilon} \left( \alpha ' - \beta ' \right) 
+ \frac{d {\rm ln} C_1}{d {\rm ln} r} \, , \nonumber
\end{eqnarray}
where ln$(r/r_{-1})$ appears since we make the expansion around $r_{-1}$.  
The first coefficient, $C_1$, is the one determining
the most shallow slope, and we find that one has $C_1 =0$ when
\begin{equation}
\alpha = \frac{\epsilon}{2} + 1 - \frac{\epsilon -2}{2\epsilon} \, 
{\rm ln} (r/r_{-1})
\, \left( \alpha ' - \beta ' (1+ \epsilon/2 )\right)
\end{equation}
which through eq.~(\ref{eq:beta}) implies
\begin{equation}
\beta_{\rm min} = 1  + \frac{1}{\epsilon} \, {\rm ln} (r/r_{-1}) \,
\left( \alpha ' - \beta ' \left( 1+ \epsilon/2 \right) \right) \, .
\label{eq:betaexpand}
\end{equation}
One sees that the density slope within radius $r_{-1}$ can be slightly
more shallow than $\beta = 1$. This is in agreement with the numerical
findings of Taylor \& Navarro~\shortcite{taylor01}.  

The most recent
simulations indeed seem to indicate that the phase-space-like density
profile, eq.~(\ref{eq:phasespace}), can indeed be well fitted with a
power law in the central resolved region, where $\epsilon$ is found to
be in the range $\epsilon \approx 2-3$ (Diemand, private
communication).  One can therefore always choose the epsilon in such a
way that $\alpha ' =0$ locally.
Furthermore, it is interesting to make a comparison with a recent
beautiful fitting formula valid for the entire resolved region, as presented
in Navarro et al.~\shortcite{navarro}
\begin{equation}
\beta _N(r) \equiv - \frac{d {\rm ln}\rho}{d {\rm ln} r} = \left( 
\frac{r}{r_{-1}} \right) ^{0.17}  \, ,
\label{eq:navarro}
\end{equation}
where $r_{-1}$ is the radius where $\beta =1$.  Since this formula
gives $\beta$ smaller than 1 it may lead to a constraint on
$\epsilon$ from the phase-space-like density.  If this formula,
eq.~(\ref{eq:navarro}), is consistent with the spherical and isotropic
Jeans equation, then this $\beta_N$ must be larger than the smallest
allowed $\beta$ as determined from eq.~(\ref{eq:betaexpand}), in the
range where the Taylor expansion leading to eq.~(\ref{eq:betaexpand})
is valid, i.e. in the vicinity inside $r_{-1}$. This is solved by
\begin{equation}
\epsilon \geq 2 \, .
\label{eq:eps}
\end{equation}
When the numerical N-body simulations reach the level of resolution
where they can resolve inside $r_{-1}$, it will be straightforward to
test the validity of the Jeans equation through the phase-space-like
density profile.  Thus, if one finds numerically that the
phase-space-like density is indeed a power law, and {\em only} so with
$\epsilon < 2$, while simultaneously the density profile is
sufficiently close to a power law (to assure validity of the Taylor
expansion, as quantified after eq.~(\ref{eq:phasespace})), then either
that region is not resolved numerically, or the
formula~(\ref{eq:navarro}) breaks down. Clearly, it is possible that
the simulations will find no density slope more shallow than $\beta
=1$, in which case there is no constraint on $\epsilon$.

\section{Conclusions}
Recent numerical dark matter simulations show that the
phase-space-{\em like} density profile, $\rho/\sigma^\epsilon$, is
well fitted locally with a simple power law with $\epsilon$ of the order
2-3.  We show that when the radial density profile is an exact power
law, $\rho \sim r^{-\beta}$, the spherically symmetric and isotropic Jeans
equations only allow the solutions where the density power slope is in
the range, $1 \leq \beta \leq 3$.  This result is independent of the value
of $\epsilon$, and shows that if the central density indeed is a power
law, then the density profile cannot be more shallow than $\beta =1$.

This constraint weakens slightly for a more general density profile
where the density power slope, $\beta(r)$, is a function of radius.
The inner density profile is then allowed to be as shallow as
described in eq.~(\ref{eq:betaexpand}). Also for anisotropic systems
more shallow profiles are allowed, according to eq.~(\ref{eq:anisotropy}).

\section*{Acknowledgments}
It is a pleasure to thank Juerg Diemand for numerous discussions and
criticism.  I thank Lucio Mayer, Ben Moore, Joachim Stadel and James
Taylor for comments and discussions.  The author thanks the Tomalla
foundation for financial support.

\label{lastpage}

\end{document}